# Enhanced spin–orbit torque *via* interface engineering in Pt/CoFeB/MgO heterostructures

Hae-Yeon Lee[1,*], Sanghoon Kim[2,3,*], June-Young Park[1], Young-Wan Oh[1], Seung-Young Park[4], Wooseung Ham[2], Yoshinori Kotani[5], Tetsuya Nakamura[5], Motohiro Suzuki[5], Teruo Ono[2,6], Kyung-Jin Lee[7,8], Byong-Guk Park[1,†]

[1]*Department of Materials Science and Engineering, KI for the Nanocentury, KAIST, Daejeon 34141, Korea*

[2]*Institute for Chemical Research, Kyoto University, Uji, Kyoto 611-0011, Japan*

[3]*Department of Physics, University of Ulsan, Ulsan 44610, Korea*

[4]*Spin Engineering Physics Team, Division of Scientific Instrument, KBSI, Daejeon 34133, Korea*

[5]*Japan Synchrotron Radiation Research Institute (JASRI), Sayo, Hyogo 679-5198, Japan*

[6]*Center for Spintronics Research Network (CSRN), Graduate School of Engineering Science, Osaka University, Osaka 560-8531, Japan*

[7]*Department of Materials Science and Engineering, Korea University, Seoul 02841, Korea*

[8]*KU-KIST Graduate School of Converging Science and Technology, Korea University, Seoul 02841, Korea*

\* These authors equally contributed to this work.

† Correspondence email: bgpark@kaist.ac.kr (B.-G. Park)

**Abstract**

Spin–orbit torque facilitates efficient magnetization switching *via* an in-plane current in perpendicularly magnetized heavy-metal/ferromagnet heterostructures. The efficiency of spin–orbit-torque-induced switching is determined by the charge-to-spin conversion arising from either bulk or interfacial spin–orbit interactions, or both. Here, we demonstrate that the spin–orbit torque and the resultant switching efficiency in Pt/CoFeB systems are significantly enhanced by an interfacial modification involving Ti insertion between the Pt and CoFeB layers. Spin pumping and X-ray magnetic circular dichroism experiments reveal that this enhancement is due to an additional interface-generated spin current of the non-magnetic interface and/or improved spin transparency achieved by suppressing the proximity-induced moment in the Pt layer. Our results demonstrate that interface engineering affords an effective approach to improve spin–orbit torque and thereby magnetization switching efficiency.

# I. Introduction

 The electrical manipulation of magnetization in magnetic nanostructures has opened up new avenues for the further development of spintronic devices because this approach affords simple device miniaturization and the potential for large-scale integration [1-3]. Conventionally, spin-transfer torque (STT) has been used to control the magnetization in magnetic multilayer structures [3,4], particularly for STT-magnetic random access memory (STT-MRAM), in which a spin-polarized current is injected in the direction perpendicular to the film plane. Recently, another type of spin-torque realized by spin–orbit coupling, the so-called spin–orbit torque (SOT) [5,6], has been widely investigated because it allows for efficient manipulation of the magnetization using in-plane current, particularly during magnetization switching [5,6] and domain-wall motion [7-9].

 SOT in heavy metal (HM)/ferromagnet (FM)/oxide heterostructures arises from the spin current induced by a charge current *via* the spin Hall effect (SHE) in the HM and/or the interfacial spin–orbit coupling (ISOC) effect at HM/FM interfaces. This spin current exerts a torque $T$ on the local magnetization as $T = \tau_{\mathrm{DL}}\hat{m} \times (\hat{y} \times \hat{m}) + \tau_{\mathrm{FL}}\hat{m} \times \hat{y}$ [10,11], where $\tau_{\mathrm{DL}}$ ($\tau_{\mathrm{FL}}$) denotes the damping-like torque (field-like torque), $\hat{m}$ the unit vector along the magnetization direction, and $\hat{y}$ the unit vector along the direction perpendicular to both directions of charge current ($\hat{x}$) and inversion symmetry breaking ($\hat{z}$). As $\tau_{\mathrm{DL}}$ which governs SOT-induced magnetization switching is known to arise mainly due to SHE in HMs [6,12], most studies have focused on finding HM materials with a large spin Hall angle $\theta_{SH}$ [13-23] for the realization of energy-efficient SOT-based spintronic devices. On the other hand, it has recently been reported that the HM/FM interface also strongly influences $\tau_{\mathrm{DL}}$ [24-32]. There are three examples; first, the magnitude and the sign of $\tau_{\mathrm{DL}}$ are changed by interface modifications [25,26], which is attributed to the contribution of the ISOC effect to $\tau_{\text{DL}}$. Second, the $\theta_{SH}$ of the HM/FM bilayer strongly depends on the spin memory loss [27] or the spin transparency [28] of its interface, thereby indicating that material engineering of HM/FM bilayers can improve the SOT efficiency. Third, the interface itself generates spin current and thus contributes to the SOT [30,31]. These results suggest that interface engineering could enable further enhancement in the SOT-induced magnetization switching efficiency.

Among various HM/FM bilayers, Pt/CoFeB is a promising candidate for spintronic device application for the following reasons: Pt is a highly conductive HM [32] when compared with other HMs such as $\beta$-W [33] or $\beta$-Ta [34], thus reducing Joule heating and thereby power consumption [32]. Moreover, CoFeB is a widely used FM in spintronic devices because of its large spin polarization in conjunction with crystalline MgO [35]. Further, a large tunnel magnetoresistance of up to 600% at room temperature has been achieved in CoFeB/MgO-based tunnel junctions [36]. However, the $\theta_{SH}$ of the CoFeB/Pt bilayer of 0.07 [24] is significantly smaller than the highest reported value among Pt/FM bilayers: ~0.2 for Pt/Co samples [28]; thus, it is of interest to further enhance the spin Hall angle.

In this study, we report significant SOT enhancement in Pt/CoFeB/MgO structures via modification of the Pt/CoFeB interface with the insertion of a thin Ti layer. We observe that the SOT-induced effective fields and magnetization switching efficiency are doubled upon introducing a 1-nm-thick Ti layer at the Pt/CoFeB interface. The enhancement in SOT or effective $\theta_{SH}$ is confirmed by means of ferromagnetic resonance (FMR) spin-pumping experiments, in which the inverse SHE of the Pt/CoFeB structure increases by twofold upon Ti insertion. Furthermore, we find that the Ti interfacial layer markedly reduces the magnetic damping constant and the Pt proximity effect. Our results demonstrate the existence of a significant interfacial contribution and suggest that interface engineering can form an efficient approach to improve SOT-induced magnetization switching.

**II. Experiment and Method**

Samples of Pt(5 nm)/Ti($t_{Ti}$)/ Co$_{32}$Fe$_{48}$B$_{20}$ (CoFeB)/MgO(1.6 nm) structures were prepared by magnetron sputtering on thermally oxidized Si substrates with a base pressure of <4.0 × 10$^{-6}$ Pa (3.0 × 10$^{-8}$ Torr) at room temperature. The Ti thickness $t_{Ti}$ was varied from 0 nm to 3 nm, and the CoFeB thickness used in this study ranged from 0.9 to 1 nm, which range guarantees strong PMA. Samples were annealed at 150 °C for 40 min under vacuum conditions to induce the PMA. The Hall bar structures with width of 5 μm were fabricated by photolithography followed by Ar-ion-beam etching. All electrical measurements were carried out at room temperature.

For the spin-pumping measurements, the barbell-shaped samples were placed on a coplanar waveguide which generated an alternating magnetic field with a frequency ranging from 8 GHz to 14 GHz. The inverse spin Hall voltage $V_{ISHE}$ was measured as a function of the external magnetic field applied along the sample plane and normal to the barbell shape [37].

The soft XAS at both the Co and Fe $L$ edges were measured using the total electron yield method with the application of 96% circularly polarized incident X-rays under an applied magnetic field of 1.9 T. The X-ray propagation direction was parallel to the film normal and the 10°-tilted magnetic field. The measurements were performed at the BL25SU beam-line at SPring8. Detailed experimental information is available in the literature (including hard XMCD measurements) [38,39]. The hard XMCD measurements were performed at the BL39XU beam-line SPring8. A transmission-type diamond X-ray phase retarder with thickness of 1.4 mm was used to achieve a high degree of circular polarization (>95%) of X-rays. The X-ray fluorescence yield mode was used to record the X-ray absorption spectra under a magnetic field of 1.9 T. The X-ray propagation direction was parallel to the film normal and the magnetic field. During scanning around the Pt $L_3$ edge, the helicity of the X-rays was reversed at 0.5 Hz.

**III. Result and Discussion**

**A. Enhancement of SOT with Ti insertion**

We firstly study the effect of the interfacial modification upon Ti insertion on SOT-induced magnetization switching using Pt(5 nm)/Ti($t_{Ti}$)/CoFeB(1 nm)/MgO Hall bar structures, where $t_{Ti}$ is varied from 0 to 3 nm (Fig. 1a). Here, we remark that Ti has a weak spin–orbit coupling [36,40] and that all films considered in this study have perpendicular magnetic anisotropy (PMA). To perform SOT-induced switching experiments, we sweep an in-plane current pulse with duration of 10 μs while measuring the anomalous Hall resistance $R_H$ between each pulse to detect the magnetization direction. During the experiment, a magnetic field of 10 mT is applied along the current (+$x$) direction to achieve deterministic switching [5,41]. Figure 1b shows the SOT-induced switching results, from which we can infer two points of note. One is the same switching polarity for all samples irrespective of $t_{Ti}$: under a positive magnetic field, positive (negative) current favors downward (upward) magnetization. This switching polarity corresponds to the positive $\theta_{SH}$ of Pt [5,41], thereby indicating that the Ti insertion layer does not affect the sign of the $\theta_{SH}$. The second point is that the critical switching current $I_C$ reduces by half when 1 nm of Ti is inserted at the Pt/CoFeB interface. On the other hand, $I_C$ becomes larger for samples with a thicker $t_{Ti}$, which is attributed to increase in current shunting through the Ti layer with a small $\theta_{SH}$. Figure 1c depicts the reciprocals of the critical switching current ($1/I_C$) and critical current density normalized by the anisotropy field ($J_C/B_k$)$^{-1}$ versus $t_{Ti}$, from which we can confirm that the switching efficiency improves in the regime where $t_{Ti}$ lies between 0.8 and 1.5 nm.

In order to quantify the effect of the Ti insertion on the SOT, we measure the SOT-induced effective magnetic fields of Pt/Ti($t_{Ti}$)/CoFeB/MgO structures using the harmonic lock-in technique [11,12]. As illustrated in Fig. 2a, the application of an ac current with a frequency of $\omega$ generates the first-harmonic Hall voltage ($V^{1\omega}$), which indicates the z-component of magnetization ($M_z$), and the second-harmonic Hall voltage ($V^{2\omega}$), which represents the oscillation of magnetization ($\Delta M_z$) due to SOT-induced effective fields: damping-like ($\Delta B_{DL}$) and field-like ($\Delta B_{FL}$) effective fields. We obtain voltages $V_x^{1\omega,2\omega}$ and $V_y^{1\omega,2\omega}$ when a magnetic field is applied longitudinally ($B = B_x$) and transversely ($B = B_y$) with respect to the current direction, respectively. Figure 2b illustrates $V_x^{2\omega}$ (closed symbols) and $V_y^{2\omega}$ (open symbols) as functions of $B_{x,y}$ for the Pt/CoFeB/MgO structures without the Ti layer, while Figs. 2c and 2d present the $V_x^{2\omega}$ and $V_y^{2\omega}$ values, respectively, for the Pt/Ti($t_{Ti}$)/CoFeB/MgO samples with $t_{Ti}$ = 1, 2, and 3 nm. The insets depict $V^{1\omega}$, which allows us to determine the polar angle of magnetization ($\theta_M$). The SOT-induced effective fields ($\Delta B_{DL}$ and $\Delta B_{FL}$) versus $\theta_M$ calculated using the measured $V^{1\omega}$ and $V^{2\omega}$ data [12] are plotted in Figs. 2e and 2f for current density $J_e$ of $1 \times 10^8 A/cm^2$. We find that the magnitude of the effective fields and their angular variation largely depend on $t_{Ti}$. The magnitudes of both $\Delta B_{DL}$ and $\Delta B_{FL}$ in the film with $t_{Ti}$ = 1 nm are enhanced by a factor of 2 when compared with those of the film without Ti, particularly for $\theta_M < 20°$, and they significantly decrease for samples with larger values of $t_{Ti}$. This result is consistent with the trend of the switching data presented in Fig. 1.

We estimate effective spin Hall angle $\theta_{\text{SH}}^{\text{eff}}$ using the conventional spin-transfer theory: $\Delta B_{\text{DL}}(0)/J_e = (\hbar/2e) \cdot (\theta_{\text{SH}}^{\text{eff}}/M_s t_{\text{CoFeB}})$ [42,43], where $\Delta B_{\text{DL}}(0)$ represents the zeroth order of $\Delta B_{\text{DL}}$[12], $\hbar$ the reduced Planck constant, $e$ the elemental charge of an electron, $M_s$ the saturation magnetization of the CoFeB layers, and $J_e$ the current density. The estimated $\theta_{\text{SH}}^{\text{eff}}$ value of the inserted 1-nm-thick Ti film is 0.19, which is significantly larger than that of the Pt/CoFeB film without Ti (0.13). This result is consistent with the enhancement in the switching efficiency; however, it is in contrast to previous results,[24,28,44] wherein the $\theta_{\text{SH}}^{\text{eff}}$ is reported to generally decrease upon insertion of an interfacial layer of a 3*d* metal. Furthermore, the introduction of the Ti interfacial layer weakens the angular dependency of the SOT and increases the ratio of $\Delta B_{\text{FL}}$ and $\Delta B_{\text{DL}}$ ($\Delta B_{\text{FL}}/\Delta B_{\text{DL}}$). This implies a substantial modification of the SOT in Pt/CoFeB/MgO structures upon insertion of the Ti layer [24].

### B. Mechanism of the large SOT with the Ti insertion

We next examine the origin of the enhancement by performing spin-pumping experiments using ferromagnetic resonance (FMR) [41] for Pt ($t_{\text{Pt}}$)/Ti (0, 1 nm)/CoFeB/MgO films, where the Pt thickness $t_{\text{Pt}}$ ranges from 2.5 to 16 nm. As schematically illustrated in Fig. 3a, when the magnetization is in resonance, a spin current is injected from the CoFeB layer into the Pt layer, which leads to increase in the effective damping constant $\Delta\alpha_{\text{eff}}$ in the CoFeB layer along with the generation of a transverse electric voltage via inverse SHE ($V_{\text{ISHE}}$) in the Pt layer. The former is related to the total spin current dissipated in the CoFeB layer while the latter is due to the spin current injected into the Pt layer. A discrepancy between the two values, $\Delta\alpha_{\text{eff}}$ and $V_{\text{ISHE}}$, can be induced if there is generation or extinction of spin current at the interface [27]. The $V_{\text{ISHE}}$ value is measured as a function of the external magnetic field, and $\Delta\alpha_{\text{eff}}$ is obtained from the variation in the line width of measured $V_{\text{ISHE}}$ spectra as a function of the FMR frequency (see Fig. 3b). Figures 3c and 3d show the Pt thickness dependence of normalized $V_{\text{ISHE}}$ by sample resistance and $\Delta\alpha_{\text{eff}}$, respectively. When the 1-nm-thick Ti interfacial layer is introduced, $V_{\text{ISHE}}$ increases, but $\Delta\alpha_{\text{eff}}$ significantly decreases. We focus on two remarkable points of interest with regard to these results. Firstly, the doubled magnitude of $V_{\text{ISHE}}$ in the Ti-inserted samples demonstrates an increase in the spin current or $\theta_{\text{SH}}^{\text{eff}}$ of Pt by a factor of 2, which is consistent with the enhanced SOT shown in Figs. 2 and 3. Secondly, a concurrent reduction in $\Delta\alpha_{\text{eff}}$ by insertion of 1-nm-thick Ti demonstrates the significant role of the interface in the spin transmission; otherwise, $\Delta\alpha_{\text{eff}}$ should be proportional to the effective spin mixing conductance and thereby to $V_{\text{ISHE}}$ [45].

We next discuss the possible origins of the enhancement in $\theta_{\text{SH}}^{\text{eff}}$ by the interfacial modification: interface-induced spin current and modified spin transparency of the FM/HM interface. We firstly consider the modification of the ISOC effect, since the Pt/CoFeB interface is replaced by Pt/Ti and Ti/CoFeB interfaces when the Ti layer is introduced. This insertion can enhance the SOT if the ISOC effect of the Pt/CoFeB interface is of opposite sign to the bulk SHE [25] or if the newly generated interface of Pt/Ti provides a positive contribution [30]. Here, we rule out the contribution of the Ti/CoFeB interface because our previous study showed a negligible SOT in the Ti/CoFeB/MgO structure [26]. To verify the modification of the ISOC effect by the Ti layer, we carry out X-ray magnetic circular dichroism (XMCD) measurements at the Fe and Co $L_{2,3}$ edges (Fig. 4a). The orbital-to-spin magnetic moment ratio ($m_o/m_s^{\text{eff}}$) of the Pt/Ti($t_{\text{Ti}}$)/CoFeB/MgO samples, which reflects the magnitude of the ISOC effect of the HM/FM interface [46,47], is estimated using the sum rule [46-49]. We find that the ratio slightly decreases with increasing $t_{\text{Ti}}$, thus indicating that the ISOC effect cannot be the reason for the enhanced SOT, as shown in Fig. 4b. Here, we remark that in this respect, the Pt/Ti interface can be a source of spin current and the SOT [30]; however, further studies are required to clarify the spin current generation from the non-magnetic interface. Another possible origin is the improvement in the interfacial spin transparency [28]. As the proximity effect in Pt is a source of spin-current depolarization [27,45], the insertion of the Ti layer may enhance the SOT by suppressing the induced moment in Pt. Thus, we investigate the effect of the Ti interfacial layer on the induced moment in Pt using hard XMCD analysis. Figure 4c shows XMCD spectra at the Pt $L_3$ edges, wherein a clear XMCD signal is observed, thus indicating a finite magnetic moment induced in the Pt $5d$ orbit for the samples without the Ti layer; however, this signal is completely eliminated by insertion of the 1-nm Ti layer. Thus, the insertion of the Ti layer can enhance the spin transparency and resultant SOT. The removal of the proximity-induced moment can also explain the reduction in $\Delta\alpha_{\text{eff}}$ upon insertion of the Ti interfacial layer (Fig. 3d) [50]. This is also supported by results indicating that the SOT monotonously decays in Ta/Ti($t_{\text{Ti}}$)/CoFeB/MgO structures with increase in Ti interfacial layer thickness, where the proximity effect in Ta is negligible (Fig. 4d).

### IV. Conclusion

In summary, we demonstrated a large enhancement in the SOT and SOT-induced switching efficiency in a Pt/CoFeB system by means of interfacial modification involving Ti insertion. The FMR spin-pumping experiments together with XMCD investigations revealed that the enhancement can be attributed to an additional interface-generated spin current and/or improved spin transparency arising due to suppression of the induced moment in Pt layer. While a quantitative analysis of the effect of the interfacial modification on the SOT requires further experimental and theoretical studies, our experimental results suggest that interface engineering is a promising approach to boost the efficiency of current-induced switching in SOT-based spintronic devices.

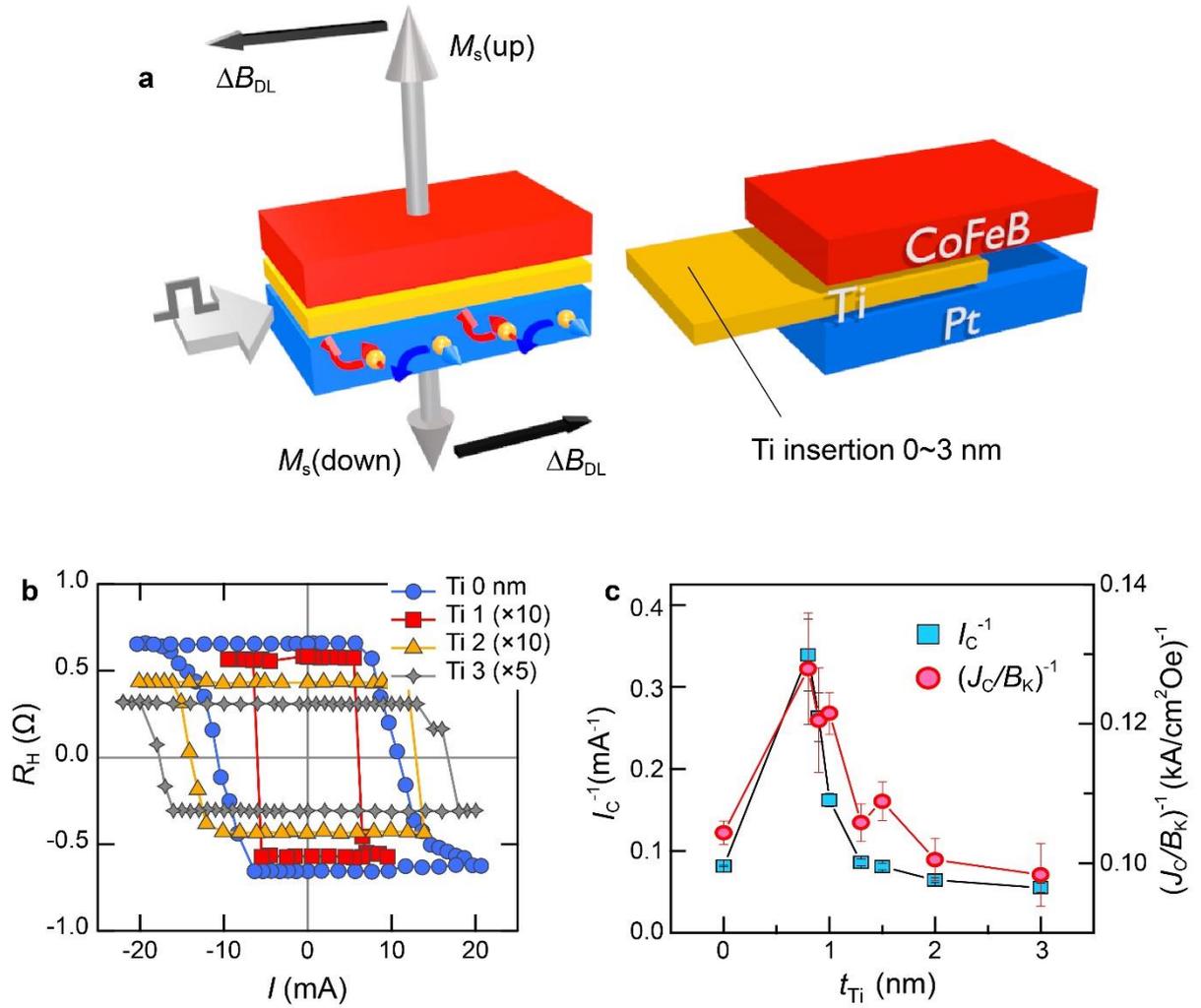

FIG. 1. (a) Schematics of spin–orbit torque (SOT)-driven magnetization switching (left) and the film structure (right). (b) The current-induced switching curves, anomalous Hall resistance $R_H$ vs. in-plane current $I$, according to the Ti thickness in Pt/Ti($t_{Ti}$)/CoFeB/MgO(1.6 nm) structures. The values of $R_H$ are normalized for comparison ($R_{H,Ti(0)} = 10R_{H,Ti(1,2)} = 5R_{H,Ti(3)}$). (c) The switching efficiency of Pt/Ti($t_{Ti}$)/CoFeB/MgO(3.2 nm) structures in terms of the reciprocal of $I_C$ (left) and $J_C/B_K$ (right). The error bars represent the standard deviations of the values obtained from three different samples.

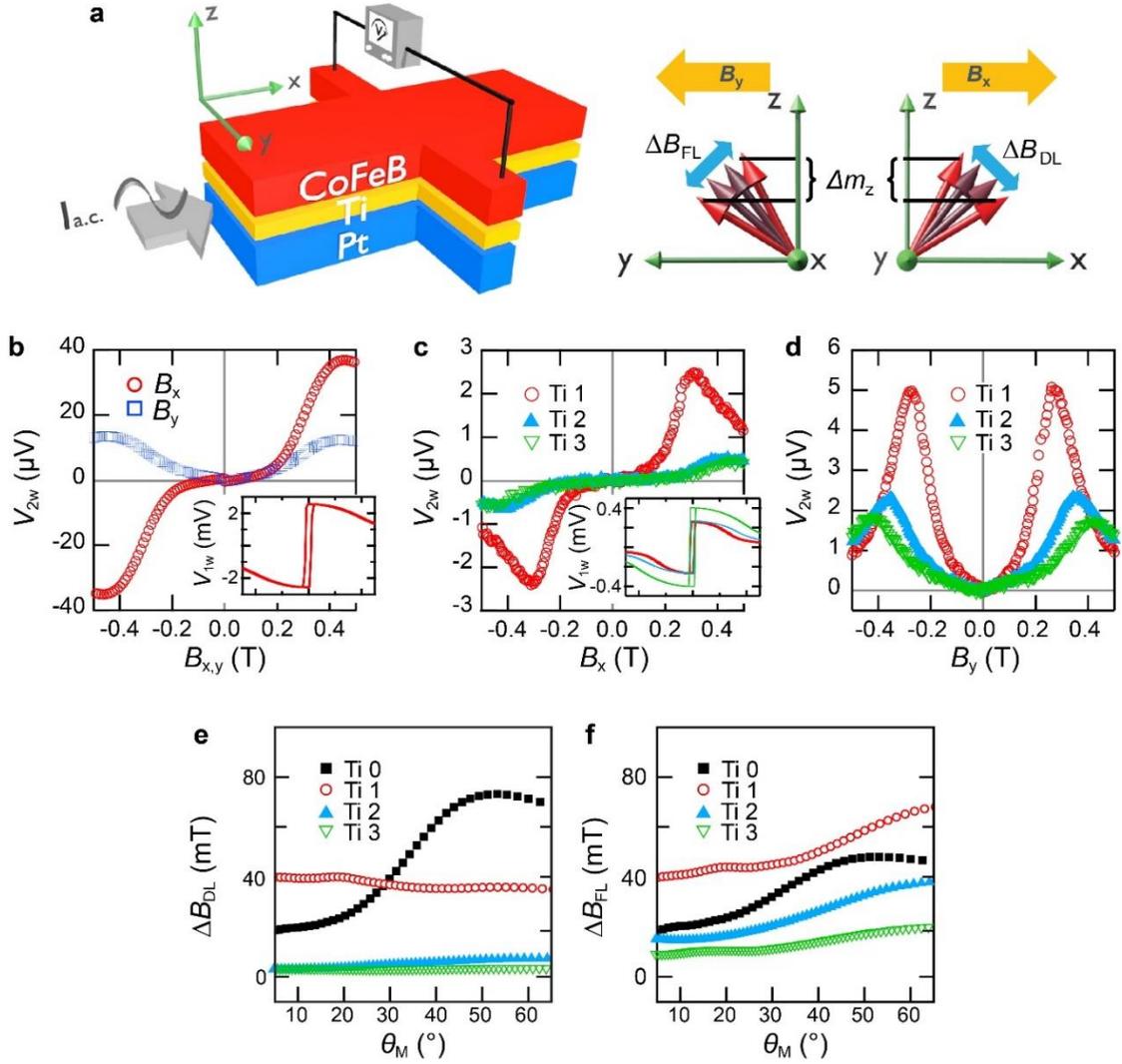

FIG. 2. (a) Schematic of measurement (left) and magnetization oscillation by $\Delta B_{\mathrm{DL}}$ and $\Delta B_{\mathrm{FL}}$ upon applying an ac current (right) (b) Second-harmonic Hall voltages $V_x^{2\omega}$ (solid symbols), $V_y^{2\omega}$ (open symbols) for the Pt/CoFeB/MgO sample. (c, d) Second-harmonic Hall voltages $V_x^{2\omega}$ (c) and $V_y^{2\omega}$ (d) for Pt/Ti($t_{\mathrm{Ti}}$)/CoFeB/MgO samples with different $t_{\mathrm{Ti}}$ values. The insets in (b) and (c) depict the first-harmonic Hall voltages $V^{1\omega}$. (e, f) Damping-like effective field $\Delta B_{\mathrm{DL}}$ (e) and field-like effective field $\Delta B_{\mathrm{FL}}$ (f) as functions of the polar angle of magnetization $\theta_M$.

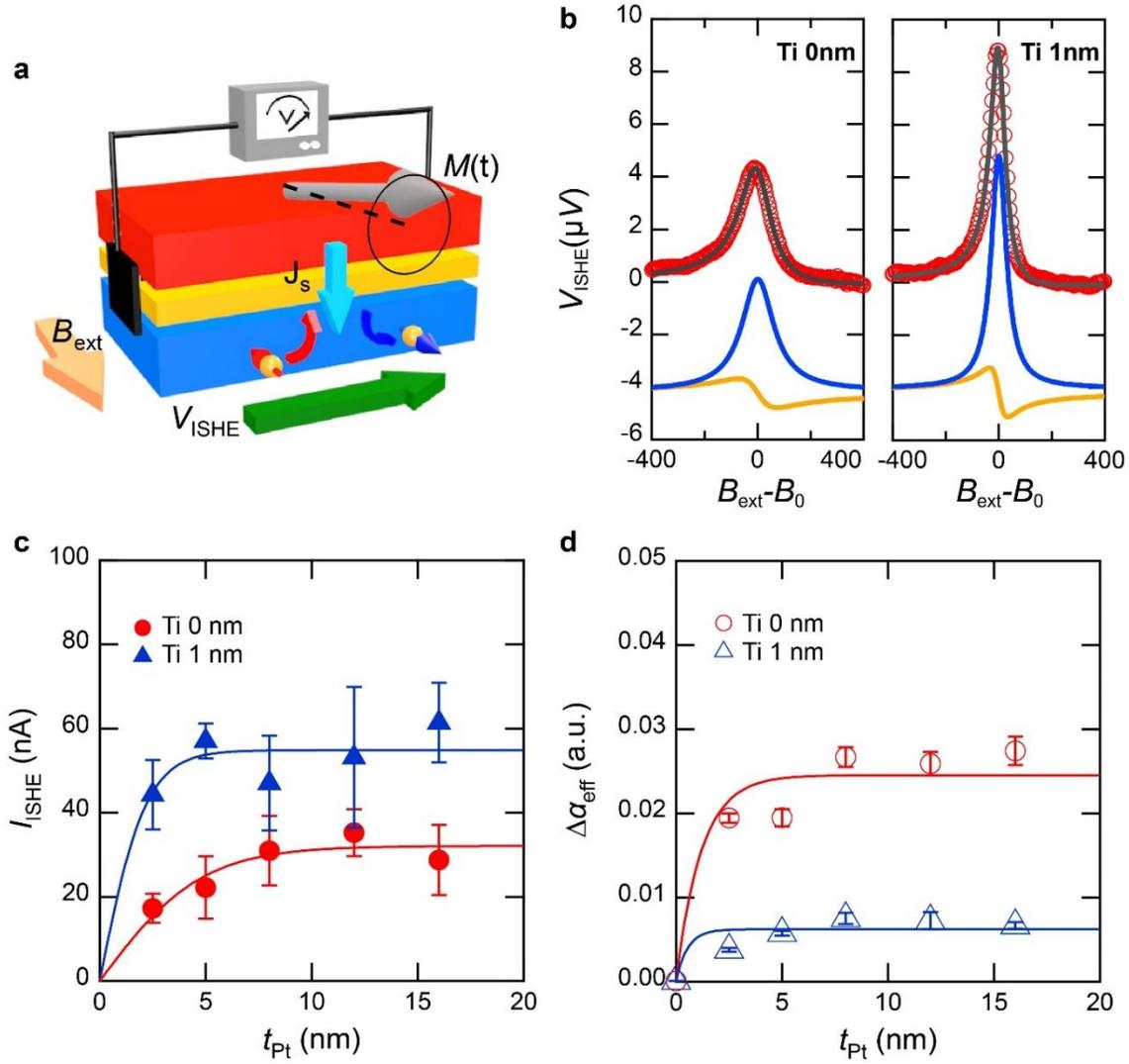

FIG. 3. (a) Schematics of the measurement. (b) Representative $V_{ISHE}$ values in Pt(12 nm)/Ti(0, 1 nm)/CoFeB(2 nm) structures. Symbols indicate measured data, while lines indicate fitting curves. The blue and yellow lines denote the decomposition of symmetric and anti-symmetric components, respectively. (c) $V_{ISHE}$ normalized by sample resistance ($I_{ISHE}$) vs. Pt thickness $t_{Pt}$ for Pt ($t_{Pt}$)/Ti(0, 1 nm)/CoFeB/MgO samples. (d) The effective damping constant $\Delta\alpha_{eff}$ vs. $t_{Pt}$ for Pt ($t_{Pt}$)/Ti(0, 1 nm)/CoFeB/MgO samples. Parameter $t_{Pt}$ varies from 2.5 to 16 nm. The lines corresponding to $V_{ISHE}$ and $\Delta\alpha_{eff}$ serve as visual guidelines.

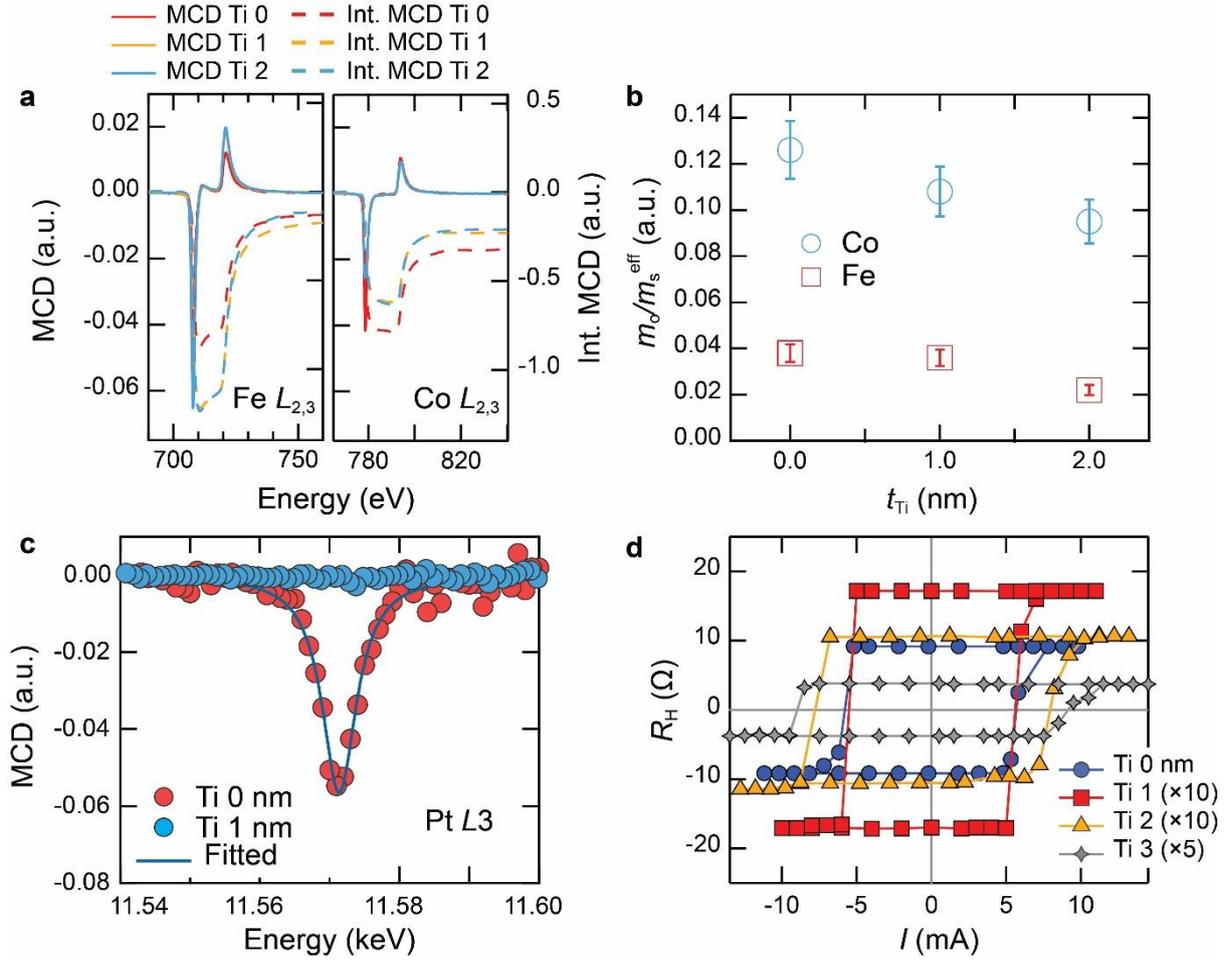

FIG. 4. (a) XMCD and integrated (int.) XMCD spectra at the Fe and Co $L_{2,3}$ edges in the Pt/Ti (0, 1, 2 nm)/CoFeB films. (b) The orbital-to-spin magnetic moment ratio ($m_o/m_s^{eff}$) values as a function of $t_{Ti}$. (c) Pt $L_3$ XMCD spectra of the Pt/ Ti (0, 1 nm)/CoFeB/MgO films. (d) Current-induced switching curves as a function of $t_{Ti}$ for Ta/Ti($t_{Ti}$)/CoFeB/MgO structures, where $t_{Ti}$ = 0–3 nm.